\documentclass{article}
\usepackage{arxiv}

\usepackage{graphicx}
\usepackage{epstopdf}
\usepackage[scaled=.97]{helvet} 
\usepackage{fancyhdr}
\usepackage{natbib}
\usepackage{url}
\usepackage{xcolor}
\usepackage{indentfirst}
\usepackage{multicol}
\usepackage{ifthen}
\usepackage{rotating}

\usepackage{amsmath}
\usepackage{amsfonts}
\usepackage{amssymb}
\usepackage{bm}
\usepackage{xcolor}
\usepackage{lmodern}
\usepackage{siunitx}

\def\pct{\%}
\renewcommand{\Pr}{\ensuremath{\operatorname{prob}}}
\def\p{\boldsymbol\pi}
\def\f{\mathbf f}
\def\x{\chi}

\NewDocumentCommand\angRange{O{} m m}{\SIrange[parse-numbers=false, #1]{\ang[parse-numbers=true]{#2}}{\ang[parse-numbers=true]{#3}}{}}

\title{Transition paths of North Atlantic Deep Water}

\author{
  Philippe Miron\\
  Center for Ocean-Atmospheric Prediction Studies\\
  Florida State University\\
  Tallahassee, FL, USA\\
   \And
  F.\ J.\ Beron-Vera\thanks{Email address for correspondence: fberon@miami.edu}\\
  Department of Atmospheric Sciences\\
  Rosenstiel School of Marine and Atmospheric Science\\
  University of Miami, Miami, FL, USA\\
  \And
  M.\ J.\ Olascoaga\\
  Department of Ocean Sciences\\
  Rosenstiel School of Marine and Atmospheric Science\\
  University of Miami, Miami, FL, USA\\
}

\begin{document}

\maketitle

\begin{abstract}
Recently introduced in oceanography to interpret the near
surface circulation, \emph{Transition Path Theory} (\emph{TPT}) is
a methodology that rigorously characterizes ensembles of trajectory
pieces flowing out from a source last and into a target next, i.e.,
those that most productively contribute to transport.  Here we use
TPT to frame, in a statistically more robust fashion than earlier
analysis, equatorward routes of North Atlantic Deep Water (NADW)
in the subpolar North Atlantic.  TPT is applied on all available
RAFOS and Argo floats in the area by means of a discretization of
the Lagrangian dynamics described by their trajectories. By considering
floats at different depths, we investigate transition paths of NADW
in its upper (UNADW) and lower (LNADW) layers. We find that the
majority of UNADW transition paths sourced in the Labrador and
southwestern Irminger Seas reach the western side of a target
arranged zonally along the southern edge of the subpolar North
Atlantic domain visited by the floats.  This is accomplished in the
form of a well-organized deep boundary current (DBC). LNADW transition
paths sourced west of the Reykjanes Ridge reveal a similar pattern,
while those sourced east of the ridge are found to hit the western
side of the target via a DBC and also several other places along
it in a less organized fashion, indicating southward flow along the
eastern and western flanks of the Mid-Atlantic Ridge.  Naked-eye
inspection of trajectories suggest generally much more diffusive
equatorward NADW routes. A source-independent dynamical decomposition
of the flow domain into analogous backward-time basins of attraction,
beyond the reach of direct inspection of trajectories, reveals a
much wider influence of the western side of the target for UNADW
than for LNADW. For UNADW, the average expected duration of the
pathways from the Labrador and Irminger Seas was found to be of 2
to 3 years. For LNADW, the duration was found to be influenced by
the Reykjanes Ridge, being as long as 8 years from the western side
of the ridge and of about 3 years on average from its eastern side.
\end{abstract}

\section{Introduction}

The North Atlantic Deep Water (NADW) formed in the Labrador and
Nordic Seas has historically been depicted as flowing equatorward,
out of the subpolar North Atlantic in the form of a well-defined
deep western boundary current (WDBC) \citep{Stommel1958}. This WDBC
constitutes the deep limb of the Atlantic Meridional Overturning
Circulation, a conduit for carbon, heat, and freshwater, acquired
at the sea surface \citep{Sabine2010}. This traditional view of the
deep circulation has been challenged by Lagrangian observations and
simulations \citep{Lozier2013, Bower2019}. The paths of NADW in its
upper layer or UNADW \citep{Lavender2000, Lavender2005, Fischer2002,
Bower2009, Bower2019, Lozier2013} and, particularly, in its lower
layer or LNADW \citep{Zou2020, Zou2017} appear to exhibit more
dispersion than originally believed.

This paper aims to scrutinize the spread of NADW more deeply than
precedent analyses via the application of \emph{Transition Path
Theory} (\emph{TPT}) \citep{E2006, VandenEijnden2006, Metzner2006,
E2010}, a methodology that has been very recently brought to
oceanography, for the interpretation of near surface circulation
\citep{Miron2021a, Drouin-etal-21}.  We will specifically apply TPT
on all available satellite-tracked profiling Argo float data and
acoustically-tracked RAFOS float data, including those collected
during the Overturning in the Subpolar North Atlantic Program (OSNAP)
\citep{Lozier2017}.  TPT will allow us to make several quantitative
assessments of the spread of NADW which lie beyond the reach of,
or are difficult to frame using, conventional Lagrangian ocean
analysis tools. These include:
\begin{enumerate}
  \item quantifying the extent by which UNADW and LNADW flow
  equatorward in the form of WDBCs;
  \item estimating the averaged time taken by UNADW and LNADW to
  exit the subpolar North Atlantic from any location within; and
  \item dynamically decomposing the flow domain into analogous
  backward-time basins of attraction for UNADW and LNADW.
\end{enumerate}

This is all possible because TPT seeks to frame the tracer trajectories
that statistically most effectively contribute to the transport
from one region of the flow domain, or \emph{source}, to another
region, or \emph{target}. Such \emph{transition paths} are formed
by trajectory pieces that connect source and target in such a way
that \emph{each trajectory piece comes out of the source last and
goes to the target next}. In other words, TPT unveils from typically
highly convoluted trajectories those portions that most productively
contribute to transport. More specifically, TPT provides rigorous
means for expressing various statistics of the ensemble of transition
pathways. These include:
\begin{enumerate}
  \item the bottlenecks during the transitions;
  \item the most likely transition channels;
  \item the rate of reactive trajectories leaving a source or
  entering a target; and 
  \item the mean duration of reactive trajectories. 
\end{enumerate}

The modeling framework for TPT analysis is provided by an autonomous,
discrete-time Markov chain \citep{Bremaud1975}, which has been
successfully used to investigate long-time asymptotics in Lagrangian
ocean dynamics \citep{Maximenko2012, Froyland2014, Miron2017,
Olascoaga2018, Beron2020, Miron2019a, Miron2019b, Miron2021a,
Drouin-etal-21}.  By combining available short-run trajectories
pieces, time-homogeneous Markov chain modeling is particularly
useful when dealing with observed trajectory records, which are of
finite-time nature.  In the present case, in particular, a limited
amount of float trajectories connect the various source and target
sets considered to attempt a direct assessment of productive
transport.

The rest of this paper is organized as follows. We begin in
Sec.~2\ref{sec:tpt} with the mathematical setup for Markov chain
modeling, followed by a self-contained exposition of the main results
of TPT. In Sec.~2\ref{sec:data-mc} we describe the construction of
the Markov chain models proposed for the evolution of the UNADW and
LNADW components of NADW using observed float data at appropriate
depths. The choice of source and target sets for TPT analysis is
rationalized in Sec.~2\ref{sec:data-tpt}. The results from the TPT
analysis are presented in Sec.~\ref{sec:results}. These are discussed
in light of results from other analyses in Sec.~\ref{sec:discussion}.
Finally, Sec.~\ref{sec:conclusion} offers a summary and the
conclusions.

\section{Methods}\label{sec:methods}

\subsection{Transition path theory}\label{sec:tpt}

Let $\x_n$ denote random position at discrete time $nT \ge 0$,
$n\in\mathbb Z_0^+$, on a closed two-dimensional flow domain $D$
covered by nonoverlapping boxes $\{b_1,\dotsc,b_N\}$. Given
$\Pr(\x_n\in b_i)$, we assume $\Pr(\x_{n+1}\in b_j) = \sum_i P_{ij}
\Pr(\x_n\in b_i)$, where
\begin{equation}
  P_{ij} := \Pr(\x_1\in b_j\mid \x_0\in b_i)
\end{equation}
is the one-step conditional probability of transitioning between
$b_i$ and $b_j$. Note that $\smash{\sum_j P_{ij} = 1}$, so $\mathsf
P = (P_{ij}) \in \mathbb R^{N\times N}$, called a \emph{transition
matrix}, is (row) stochastic. Let $x(t)$ represent a very long float
trajectory visiting every box of the covering of $D$ many times.
In practice, there are many finite-length float trajectories that
sample $D$ well. Let $x(t)$ and $x(t+T)$ at any $t>0$ provide
observations for $\x_0$ and $\x_1$, respectively. These are used
to approximate $P_{ij}$ via transition counting, viz.,
\begin{equation}
  P_{ij} \approx \frac{C_{ij}}{\sum_k C_{ik}},\quad
  C_{ij} := \#\big\{x(t)\in b_i,\, x(t+T)\in b_j,\, t:\text{any}\big\}.
\end{equation}
\label{eq:P}%
In other words, the float motion is envisioned as that of random
walkers along an autonomous, discrete-time \emph{Markov chain}
\citep[cf., e.g.,][]{Bremaud1975}. At the continuous level, the
Lagrangian dynamics as described by the float trajectories are
assumed to obey a time-homogeneous stochastic (i.e.,
steady-advection--diffusion) process.
i
The chain will be assumed to be \emph{ergodic} (i.e., such that all
its states (boxes of the covering of $D$) communicate irrespective
of the starting state) and \emph{mixing} (i.e., such that none of
its states is revisited cyclically). In these conditions, eigenvalue
one of $\mathsf P$ is both maximal and simple. The corresponding
\emph{left} eigenvector, $\p = (\pi_i) \in \mathbb R^{1\times N}$,
can be chosen componentwise positive, and is both invariant and
limiting, i.e., $\p = \p \mathsf P = \lim_{n\uparrow\infty}\f \mathsf
P^n$ for any $\f\in \mathbb R^{1\times N}$. Normalized to a
\emph{probability} vector (i.e., so $\sum_i \pi_i = 1$), $\p$
represents a \emph{stationary distribution}, which will be assumed
to set the long-time asymptotics of $\{\x_n\}$, namely, $\Pr(\x_n\in
b_i) = \pi_i$. This will allow us to investigate generic aspects
of the float motion in statistical stationarity, rather than
particular aspects bound to initial conditions.

\begin{figure*}[ht!]
 \centering \includegraphics[width=.5\textwidth]{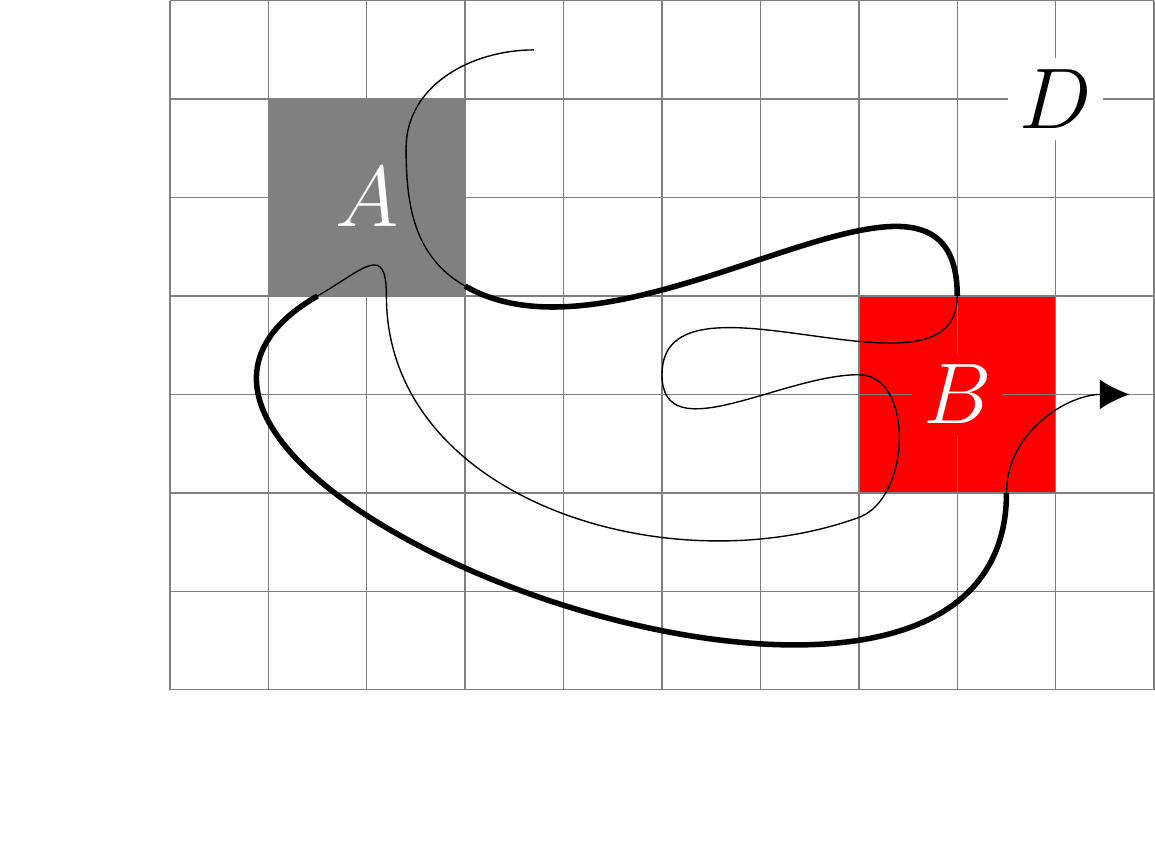} \caption{Schematic
 representation of a piece of a hypothetical infinitely long float
 trajectory (black) that densely, albeit not necessarily uniformly,
 fills a closed flow domain $D$, partitioned into boxes (black).
 Indicated in gray and red respectively are source ($A$) and target
 ($B$) sets. Highlighted by the thick (black) lines are two members
 of an ensemble of reactive trajectories. These are the trajectory
 subpieces that connect the boundary of $A$ with the boundary of
 $B$ in direct transition from $A$ to $B$, i.e., without returning
 back to $A$ or going through $B$ in between.}
\label{fig:tpt}
\end{figure*}

The TPT of \citet{E2006} provides a rigorous approach to study
transitions from a set $A\subset D$ to another, disjoint set $B\subset
D$. The pieces of trajectories running from $A$, referred to as
\emph{source}, to $B$, referred to as \emph{target}, without going
back to $A$ or going through $B$ in between, are the main focus and
known as \emph{reactive trajectories} (Fig.~\ref{fig:tpt}). This
uses traditional jargon that identifies source set $A$ with the
reactant of a chemical transformation and target set $B$ with its
product. Reactive trajectories are also known as \emph{transition
paths}; will use this terminology or similar to refer to them here
too. Most importantly, they describe pathways that \emph{contribute
most effectively to the transport from $A$ to $B$}. For the problem
of interest, transition \emph{float} paths from a set $A$ within
the subpolar gyre to a set $B$ located at its southern edge will
highlight the most effective export paths of the deep water masses
sampled by the floats.

The main objects of TPT are the \emph{forward}, $\mathbf q^+ =
(q^+_i)\in \mathbb R^{1\times N}$, and \emph{backward}, $\mathbf
q^- = (q^-_i)\in \mathbb R^{1\times N}$, \emph{committor probabilities}.
These give the probability of a random walker initially in $b_i$
to first enter $B$ and last exit $A$, respectively.  Namely, $q_i^\pm
: = \Pr(\tau^\pm_B < \tau^\pm_A \mid \x_0 \in b_i)$, where $\tau^\pm_S
:= \inf\{nT : \x_{\pm n}\in S\}$ with the plus (resp., minus) sign
denoting \emph{first entrance} (resp., \emph{last exit}) \emph{time}
of a set $S\subset D$. The committors are fully computable from
$\mathsf P$ and $\p$, since $\Pr(\x_n\in b_i) = \pi_i$, according
to:
\begin{equation}
  q^\pm_i = \sum_j P^\pm_{ij}q^\pm_j,\,i\in\overline{A\cup B},
  \quad q^\pm_{i\in A} = \delta_{1\mp1,2},\quad
  q^\pm_{i\in B} = \delta_{1\pm1,2}.
  \label{eq:q}
\end{equation}	
Here, $\mathsf P^+ = \mathsf P$, $P^-_{ij} := \Pr(\x_0=j\mid \x_1=i)
= \frac{\pi_j}{\pi_i}P_{ji}$ are the entries of the \emph{time-reversed}
transition matrix, i.e., for the original chain traversed in backward
time, $\{\x_{-n}\}$; the overbar means complement; and the short-hand
notation $i\in S$ for $i : b_i\subset S$ is herein used.

The committor probabilities are used to express several statistics
of the ensemble of reactive trajectories as follows
\citep[e.g.,][]{Metzner2006, Helfmann2020}.
\begin{enumerate}
  \item The \emph{distribution of reactive trajectories}, $\boldsymbol
  \mu^{AB} = (\mu^{AB}_i)\in \mathbb R^{1\times N}$, where $\mu^{AB}_i$
  is defined as the joint probability that a trajectory is in box
  $b_i$ while transitioning from $A$ to $B$ and is computable as
  \begin{equation}
	\mu^{AB}_i = q^-_i\pi_iq^+_i,
	\label{eq:mu}
   \end{equation}
   describes the bottlenecks during the transitions, i.e., where
   reactive trajectories spend most of their time. Clearly,
   $\mu^{AB}_{i\in A\cup B} \equiv 0$.
   \item The \emph{effective current of reactive trajectories},
   $\mathsf f^+ = (f^+_{i,j})\in \mathbb R^{N\times N}$, where
   $f^+_{i,j}$ gives the net flux of trajectories going through
   $b_i$ at time $nT$ and $b_j$ at time $(n+1)T$ on their way from
   $A$ to $B$, indicates the most likely transition channels. This
   is computable according to
   \begin{equation}
	 f^+_{ij} = \max\left\{f^{AB}_{ij} -
	 f^{AB}_{ji},0\right\},\quad
	 f^{AB}_{ij} = (1-\delta_{ij})q^-_i\pi_jP_{ij}q^+_j.
	 \label{eq:f}
   \end{equation}
   To visualize reactive trajectories, one can proceed in two
   different but complementary ways, as we do here.
   \begin{enumerate}
     \item On one hand, one can simply depict the magnitude and the
     direction of the effective reactive current $\mathsf f^+$ out
     of each box $b_i$ of the flow domain covering. This is done
     by attaching to each $b_i$ the two-dimensional vector on the
     ``plane'' $\smash{\sum_{j\neq i}f^+_{ij}\vec e_{ij}}$, where
     $\vec e_{ij} \in \mathbb R^{2\times 1}$ is the unit vector
     \emph{pointing from the center of box $b_i$ to the center of
     $b_j$} \citep{Helfmann2020, Miron2021a}.
     \item On the other hand, one can depict \emph{dominant transition
     paths}, which maximize the minimal effective current along the
     reactive trajectories \citep{Metzner2009}. The larger the
     minimal effective current along a reactive trajectory, the
     more current it conducts from $A$ to $B$. In practice one
     applies a flow decomposition algorithm that seeks to concatenate
     ``bottlenecks'' in associated $f^+$-weighted directed graphs,
     i.e., edges with minimal effective current \citep[cf.][for
     details]{Metzner2009}. Dominant transition \emph{float} paths
     between appropriately chosen source and target sets will unveil
     the major deep water mass parcel pathways connecting them most
     productively.
   \end{enumerate}
   \item The \emph{rate of reactive trajectories} leaving $A$ or
   entering $\tilde B \subseteq B$, defined respectively as the
   probability per time step of a reactive trajectory to leave $A$
   or enter $\tilde B \subseteq B$, are computed respectively as
   \begin{equation}
	 k^{A\to} = \sum_{i\in A, j} f^{AB}_{ij},\quad 
	 k^{\tilde B\leftarrow} = \sum_{i, j\in\tilde B} f^{AB}_{ij},
	 \label{eq:k}
   \end{equation}
   and can be interpreted in various ways. One is as the proportion
   of reactive trajectories leaving $A$ or entering $\tilde B
   \subseteq B$. If divided by $T$, the other possible interpretation
   is as the frequency at which a reactive trajectory leaves $A$
   or enters $\tilde B\subseteq B$. In particular, $\smash{k^{A\to}
   \equiv k^{B\leftarrow}}$. Moreover, in a fully three-dimensional
   calculation, with three-dimensional boxes rather than ``tiles''
   as considered in the present two-dimensional analysis constrained
   by the data availability, reactive rates can be interpreted as
   volumetric flow rates. More specifically, upon dividing
   $\smash{k^{A\to}}$ or $\smash{k^{\tilde B\leftarrow}}$ by $T$
   and multiplying the result by the volume spanned by the box
   covering of $D$, one would get the volumetric flow rate out of
   $A$ or into $\tilde B$, respectively.
   \item Finally, the \emph{expected duration}, $t^{AB}$, of a
   transition from $A$ to $B$ is obtained by dividing the probability
   of being reactive by the transition rate interpreted as a
   frequency, viz.,
   \begin{equation}
     t^{AB} = \frac{\sum_{j\in\overline{A\cup B}}
	 \mu^{AB}_j}{k^{AB}}.
	 \label{eq:t}
   \end{equation}
\end{enumerate}

When $D$ represents an open flow domain, i.e., with trajectories
flowing out and returning back in as is the case of this work,
$\mathsf P$ is no longer stochastic, which requires an adaptation
of TPT \citep{Miron2021a}. This involves augmenting chain $\mathsf
P$ by a stochastic transition matrix $\tilde {\mathsf P}\in \mathbb
R^ {(N+1)\times (N+1)}$ defined by
\begin{equation}
 \tilde {\mathsf P} :=
 \begin{pmatrix}
	 \mathsf P & \mathbf p^{D\to\omega}\\ 
	 \mathbf p^{D\leftarrow\omega} & \mathbf 0
 \end{pmatrix}
 \label{eq:closure}
\end{equation}
where $\omega$ is the state, called a \emph{two-way nirvana state},
used to augment the chain defined by $P$. In eq.\ \eqref{eq:closure},
vector $\smash{\mathbf p^{D\to\omega} := \big(1 - \sum_{j\in D}
P_{ij}\big)^\top\in \mathbb R^{N \times 1}}$ gives the outflow from
$D$ and the \emph{probability} vector $\smash{\mathbf
p^{D\leftarrow\omega}\in \mathbb R^{1\times N}}$ gives the inflow.
When possible, this is estimated from the trajectory data. Other
possibilities exist, for instance when there are not enough trajectory
observations or their limited length prevents them from capturing
the return flow, as is for example the case with the deepest floats
analyzed in this study. One such possibility is redistributing any
imbalance like the \emph{quasi}stationary distribution $\p$,
normalized by $\sum_i \pi_i$, or uniformly along the chain. However,
the TPT results of this paper were not found sensitive to any of
these choices. The expectation is that the dynamics produced by the
restriction of $\tilde {\mathsf P}$ to $D$, $\smash{\tilde {\mathsf
P}\vert_{D}}$, is consistent with the original dynamics, i.e.,
produced by $\mathsf P$, under the assumption of well-mixedness
between exit from $D$ and reentry into it. TPT is adapted in
\citet{Miron2021a} such that transitions between $A$ and $B$ are
constrained to take place within $D$, i.e., they avoid $\omega$.
This is accomplished by replacing $\p$ in the TPT formulae above
by $\smash{\tilde\p\vert_D}$, where $\tilde\p$ is the stationary
distribution of $\tilde {\mathsf P}$.

We finalize this section by noting that, theoretically, by ergodicity
of a Markov chain, the TPT statistics can be computed by ``counting''
transition events of an infinitely long, $\p$-distributed trajectory
\citep{VandenEijnden2006, Helfmann2020}. For instance, the forward
committor $q_i^+$ is equal to the fraction of all visits paid by
such a trajectory to box $b_i$ after having directly transitioned
to $B$ without hitting $A$ first. One may then wonder why all the
TPT sophistication is really necessary when one could simply do an
approximation by counting. The answer is in the nature of the
trajectory data: sufficiently many trajectories that are sufficiently
long and appropriately distributed would be needed to resolve the
transition statistics. None of these conditions are satisfied in
practice, not even if the trajectory data are generated numerically,
and the best available option is to combine all available trajectory
information into a Markov chain.

\subsection{Construction of the Markov chain model(s)}\label{sec:data-mc}

The Markov chain is constructed using trajectories from Argo and
RAFOS floats deployed in, or travelling through, the subpolar North
Atlantic above \ang{40}N.  In this region, the inclusion of Argo
floats is critical to the analysis of the NADW circulation due to
the insufficient density of RAFOS floats, which drift at a constant
depth. Given the nature of Argo floats, which surface from their
parking depth every ten days, we note that surface currents account
on average for 15\pct{ }of Argo float displacement. Since 2013, the
majority of Argo floats use the Iridium telecommunications service,
which allows for faster transmission, decreasing the time period
at the surface, as well as the influence of surface currents on the
float trajectory.  On the other hand, in \citet{Miron2019a} it was
shown that the vertical excursions of the Argo floats do not
substantively affect the description of the Lagrangian motion at
their parking depth.  Thus we proceed with confidence and consider
Argo and RAFOS floats together in the analysis.

\begin{figure*}[ht!]
  \centering \includegraphics[width=\textwidth]{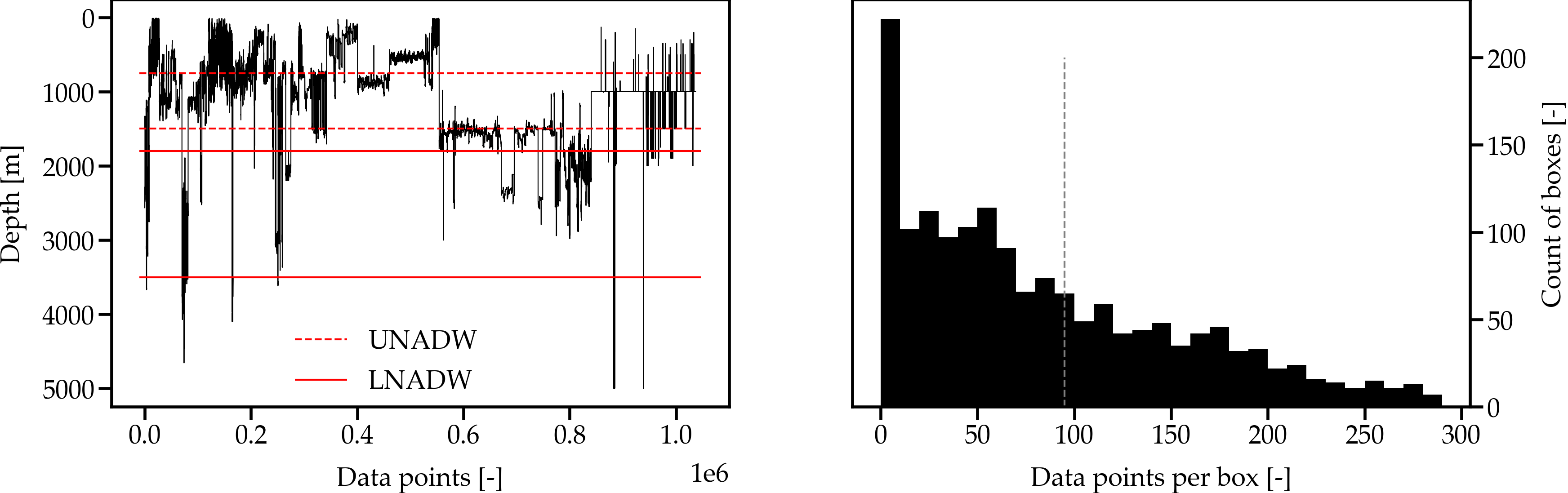} \caption{(left)
  Independent of horizontal position or time, depth reached by each
  RAFOS or Argo float deployed in, or traversing the, subpolar North
  Atlantic (Fig.~\ref{fig:maps}) since the 1980s. The broken red
  lines indicate the limits of the depth ranges chosen to construct
  the Markov chains that represent the motion of the UNADW (750--1500
  m) and LNADW (1800--3500 m) components. (right) Histogram of the
  raw data density within boxes of the domain (above 50$^\circ$N)
  with the broken gray line indicating the average value of 95 data
  points per box. Replace the right panel to include an histogram
  of the data density.}
\label{fig:dataset}
\end{figure*}

The Argo floats deployed in the Labrador Sea during the World Ocean
Circulation Experiment and the Deep Convection Experiment in the
1990s, and elsewhere since the 1980s, encompass most (1513 of a
total of 2339 floats) of the data record (Fig.~\ref{fig:dataset}).
The OSNAP program contributes to the data record over the period
2014--2017 with 134 RAFOS floats, deployed in various locations of
the subpolar North Atlantic. There are 1745 floats sampling depths
between 750 and 1500 m, and 302 floats between 1800 and 3500 m.
Observational trajectories in these two depth ranges are here used
to construct two independent Markov chains, describing the evolution
of the UNADW and LNADW components of NADW, respectively. The
construction of the two Markov chains first involves covering each
region $D$ sampled by each set of trajectories, with a grid of
$0.6^\circ \times 0.6^\circ$ boxes. Henceforth, we differentiate
between the shallow Markov chain, describing the circulation of
UNADW (Fig.~\ref{fig:maps}, left panel) and the deep Markov chain,
describing the circulation of LNADW (Fig.~\ref{fig:maps}, right
panel). The area of the boxes is not uniform due to the Earth's
sphericity, but this is of no consequence if the discretization of
the Lagrangian dynamics is carried out appropriately;
cf.~\citet{Miron2019a}. To compute the transition matrix $P$ of
each Markov chain, we use the float positions at any time $t$ and
$T=\SI{10}{days}$ later to provide observations of random positions
$\chi_0$ and $\chi_1$. The transition time ($T=\SI{10}{days}$) is
sufficiently long to guarantee negligible memory into the past (we
have estimated a Lagrangian decorrelation time of about 5 days) so
the Markov assumption can be expected to approximately hold. This
$T$ choice also guarantees sufficient communication among boxes,
while still maximizing sampling. The results presented below,
however, are insensitive to $T$ choices between \SIrange{5}{20}{days}.
Similar results were found earlier by \citet{Miron2019a} using RAFOS
and Argo float data in the Gulf of Mexico.

An important cautionary note is that UNADW and LNADW are usually
defined by their (mass) density signature rather than depth ranges.
Commonly considered sigma-potential density ranges are 27.66--27.8
and 27.8--27.88 kg m$^{-3}$, respectively \citep{Lozier2019}. These
respectively lie, typically, within the depth ranges 750--1500 m
and 1800--3500 m (discarding data between 1500--1800 m near their
interface) here used to differentiate between the two water masses
in question. Their choice had the above observation in mind, but
it was also constrained by the data availability. Thus our NADW
flow description admittedly is an approximation.

\subsection{Source and target sets for TPT analysis}\label{sec:data-tpt}

\begin{figure*}[ht!]
  \centering \includegraphics[width=\textwidth]{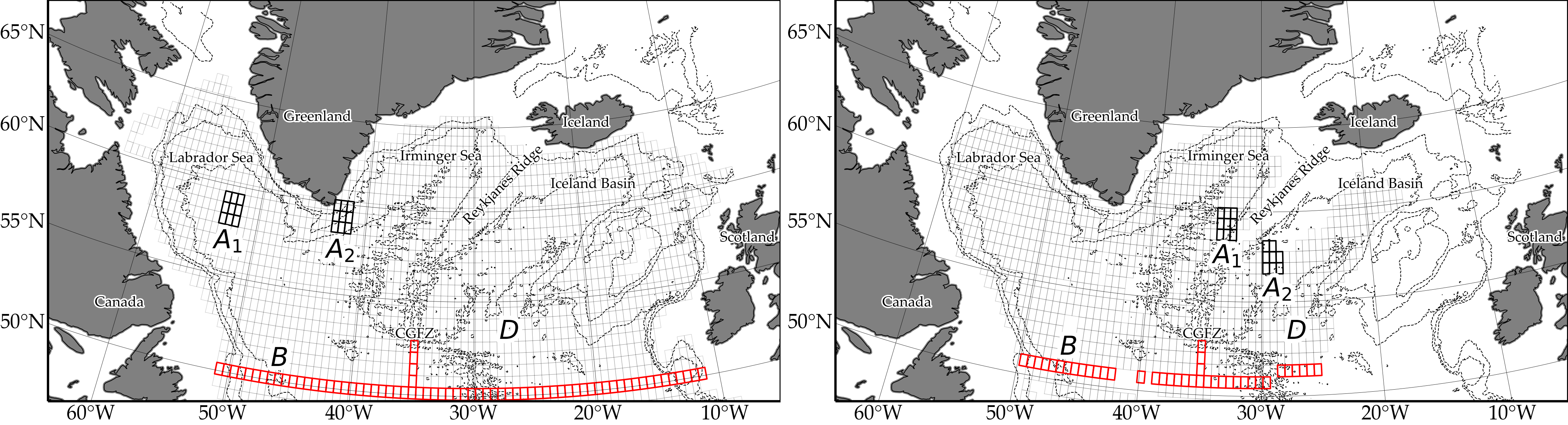} \caption{Box
  (gray) of the subpolar North Atlantic regions defining the UNADW
  (left) and LNADW (right) Markov chain models, constructed using
  float data within 750--1500 m and 1800-3500 m, respectively.
  Source and target sets for TPT analysis are indicated in black
  and red, respectively. Isobaths \SIlist{1000;2000;3000}{\meter}
  are shown in this and in all subsequent figures for reference.}
  \label{fig:maps}
\end{figure*}

Two source sets, $A_1$ and $A_2$, and a single target set $B$, given
by the union of nearly contiguous boxes, are defined for each of
the two Markov chains (Fig.~\ref{fig:maps}). For the UNADW chain
(Fig.~\ref{fig:maps}, top panel), sets $A_1$, positioned in the
middle of the Labrador Sea, and $A_2$, located in the southwestern
Irminger Sea off the southern tip of Greenland. For the LNADW chain
(Fig.~\ref{fig:maps}, bottom panel), sets $A_1$ and $A_2$ straddle
the Reykjanes Ridge. The UNADW sources in the Labrador Sea and the
southwestern Irminger Sea lie in regions of presumed deep water
formation \citep{Pickart2003}. The southwestern Irminger Sea and
the LNADW sources overlap with the location of an OSNAP mooring
array or coincide with the release sites of RAFOS floats
\citep{Ramsey2020}. In particular, the LNADW source east of the
Reykjanes Ridge can be expected to intersect Iceland--Scotland
Overflow Water. Constrained by the availability of data and the
depth of the water masses across the North Atlantic, the target set
$B$ is taken to represent the \emph{southern edge} of the subpolar
North Atlantic, so the TPT analysis enables assessing pathways of
NADW out of it (Fig.~\ref{fig:maps}). For the UNADW chain, which
has a higher density of float data, this crosses the domain at a
latitude of \ang{50}N. For the LNADW chain, for which the float
coverage is poorer, it is positioned approximately along \ang{51}N.
Note that each set $B$ includes a meridional set of boxes along
\ang{35}{N} extending out to the Charlie--Gibbs Fracture Zone (CGFZ),
at about \ang{53}N. This set of target boxes is included to assess
internal pathways in the subpolar North Atlantic. 

\section{Results}\label{sec:results}

\subsection{Transition pathways of UNADW}

\begin{figure*}[ht!]
  \centering \includegraphics[width=\textwidth]{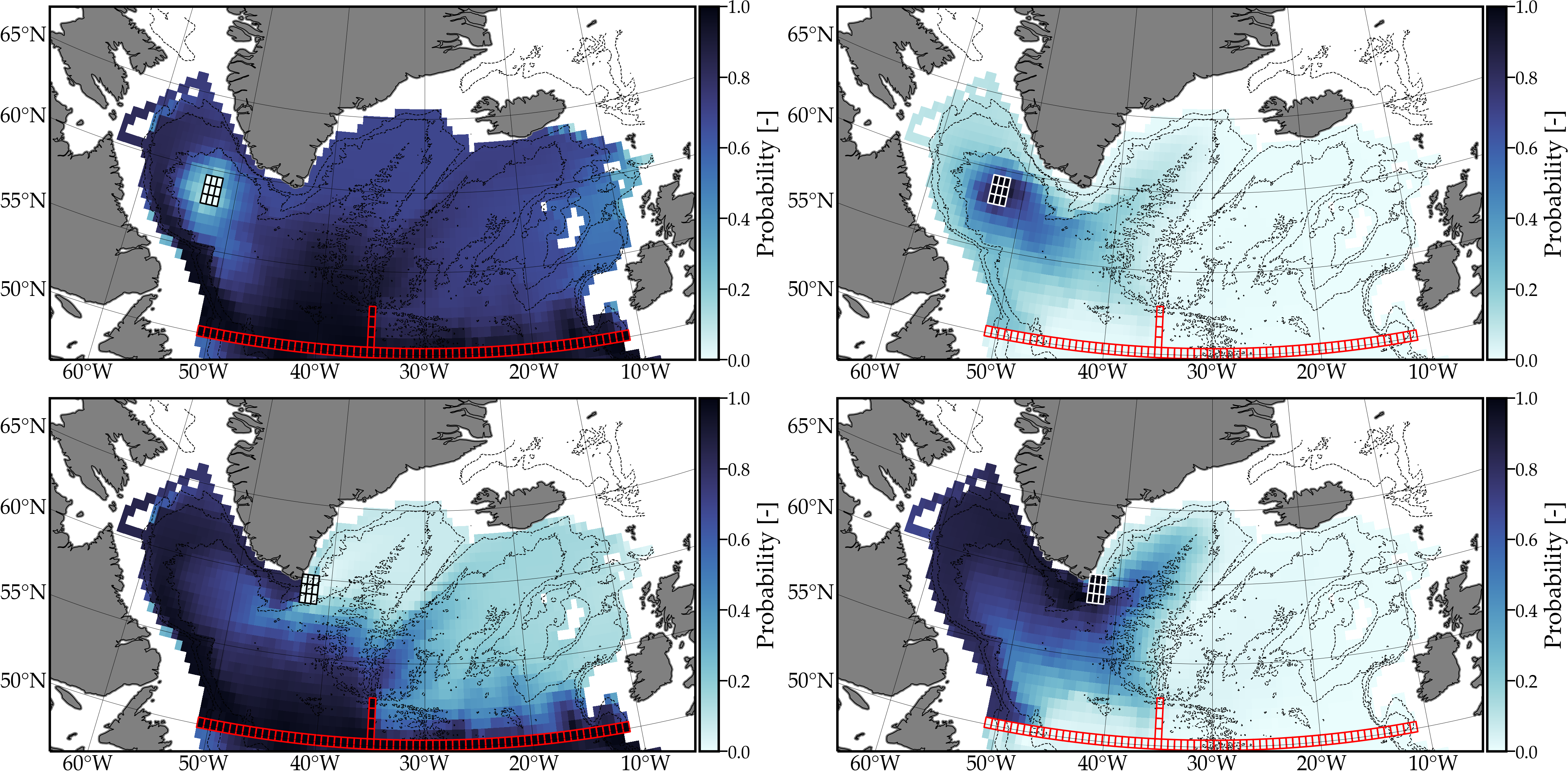} \caption{(top)
  For the UNADW Markov chain model with the source in the Labrador
  Sea, probability of trajectories to commit to the target (red
  boxes) in forward time (left) and to the source (black/white
  boxes) in backward time (right). (bottom) As in the top panels,
  but with the source in the Irminger Sea.} \label{fig:committor_lsw}
\end{figure*}

\begin{figure*}[ht!]
  \centering \includegraphics[width=\textwidth]{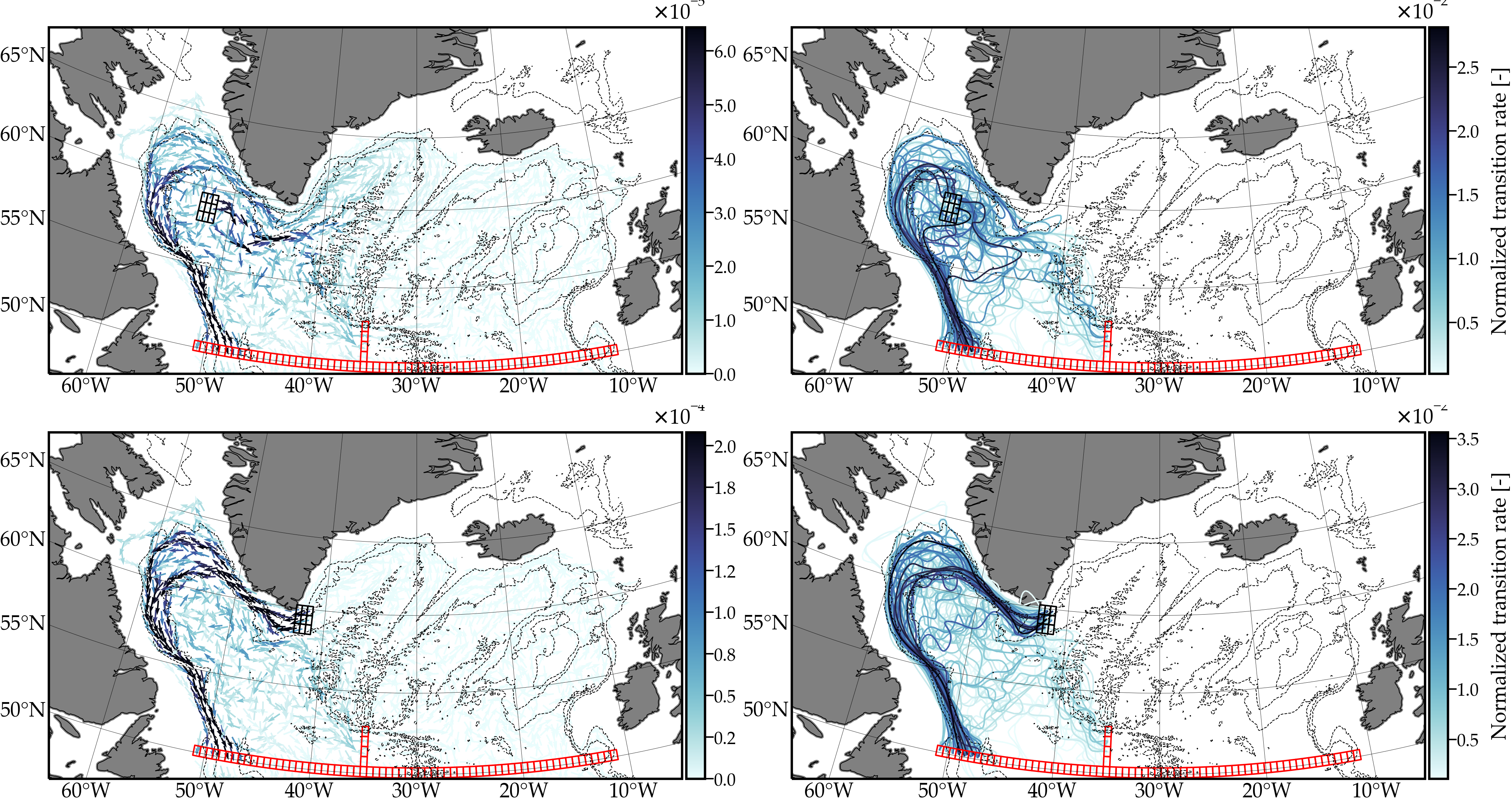} \caption{(top)
  For the UNADW chain with the source in the Labrador Sea, effective
  reactive currents (left) and dominant transition paths accounting
  for 90\pct{ }of the probability flux of reactive trajectories
  (right). (bottom) As in the top panels, but with the source in
  the Irminger Sea.} \label{fig:current_lab}
\end{figure*}

We begin by discussing the structure of the UNADW forward ($\mathbf
q^+$) and backward ($\mathbf q^-$) committor probabilities; cf.\
eq.\ \eqref{eq:q}. These are shown in the left and right panels of
Fig.~\ref{fig:committor_lsw}, with the source placed in the Labrador
(top) and southwestern Irminger (bottom) Sea. The probability of
the trajectories to commit to the target in forward time increases
toward the target. Similarly, their probability to commit in backward
time to the source increases toward the source. By construction,
the forward committor vanishes at the source and is maximal at the
target, and vice versa for the backward committor. Clearly, there
is no reactive current inside the source and the target, where
$q_i^+ = 0$ and $q_i^- = 0$ for each box $b_i$ covering these sets,
respectively, and it will be small in regions covered by boxes where
$q_i^+ \approx 0$ \emph{or} $q_i^- \approx 0$. This takes place in
most of the eastern subpolar North Atlantic (Fig.~\ref{fig:current_lab}).

More specifically, in the left and right panels of
Fig.~\ref{fig:current_lab} the resulting effective reactive currents
of UNADW are depicted alongside their associated dominant transition
paths, with the top and bottom rows corresponding to the Labrador
and Irminger Sea sources, respectively. The dominant transition
paths are colored according to the minimal effective reactive current
along their path.

As expected from the analysis of the committors, most of the reactive
transitions of UNADW take place in the western subpolar North
Atlantic. From both sources, the majority of the transition pathways
reach the target at its western edge along a well-defined DBC,
seemingly consistent with $f/h$ conservation, where $f$ is the
Coriolis frequency and $h$ is depth \citep{LaCasce2000}.

For UNADW emerging from the source centered in the Labrador Sea
(Fig.~\ref{fig:current_lab} top), a sizeable amount of reactive
current is initially directed eastward, into the Irminger Sea, where
it reaches the East Greenland Current and flow around Greenland.
Dominant pathways (top-right panels, accounting for 90\pct{ }of the
probability flux of reactive trajectories) emerging from the Labrador
Sea show some degree of isotropicity, consistent with $f/h$
conservation in a region where the bottom topography is relatively
flat. The majority of dominant paths eventually join to form a WDBC,
and swiftly reach the southwestern edge of the domain. In addition,
some interior transition paths are seen to connect the Labrador Sea
source with the meridional of boxes of the target at \ang{35}W,
indicating possible transition pathways of UNADW connecting the
western and eastern subpolar North Atlantic through the CGFZ,
previously identified by \cite{Lavender2000, Lavender2005} and more
recently quantified by \cite{Goncalves2020}. From the Labrador Sea,
such internal paths represent a small percentage, not exceeding
10\pct, of the total amount of transitions between the Labrador Sea
source and the target.

For UNADW emerging from the southwestern Irminger Sea source
(Fig.~\ref{fig:current_lab} bottom), the DBC is restricted between
the 1000- and 3000-m isobaths as it progress northwestward bordering
western Greenland. At about \ang{61}N where the 2000- and 3000-m
isobaths separate from one another, the reactive current splits
into two branches, one roughly following the 3000-m isobath and
another one following the 2000-m isobath very closely. The two
branches eventually merge, roughly off the northeastern end of the
Labrador Peninsula, where the 2000- and 3000-m isobaths get closer
again, into a single-branch DBC that flows out of the subpolar North
Atlantic. The two UNADW paths inferred by the TPT analysis are hard
to reveal from the inspection of individual float trajectories. The
two DBC branches unveiled by TPT provide observational support to
inferences made from the analysis of simulated float trajectories
\citep{Jong2016} as well as robust statistical confirmation of
earlier inferences made from the inspection of a subset of the float
data considered here \citep{Cuny2002}.

\subsection{Transition pathways of LNADW}

\begin{figure*}[ht!]
  \centering \includegraphics[width=\textwidth]{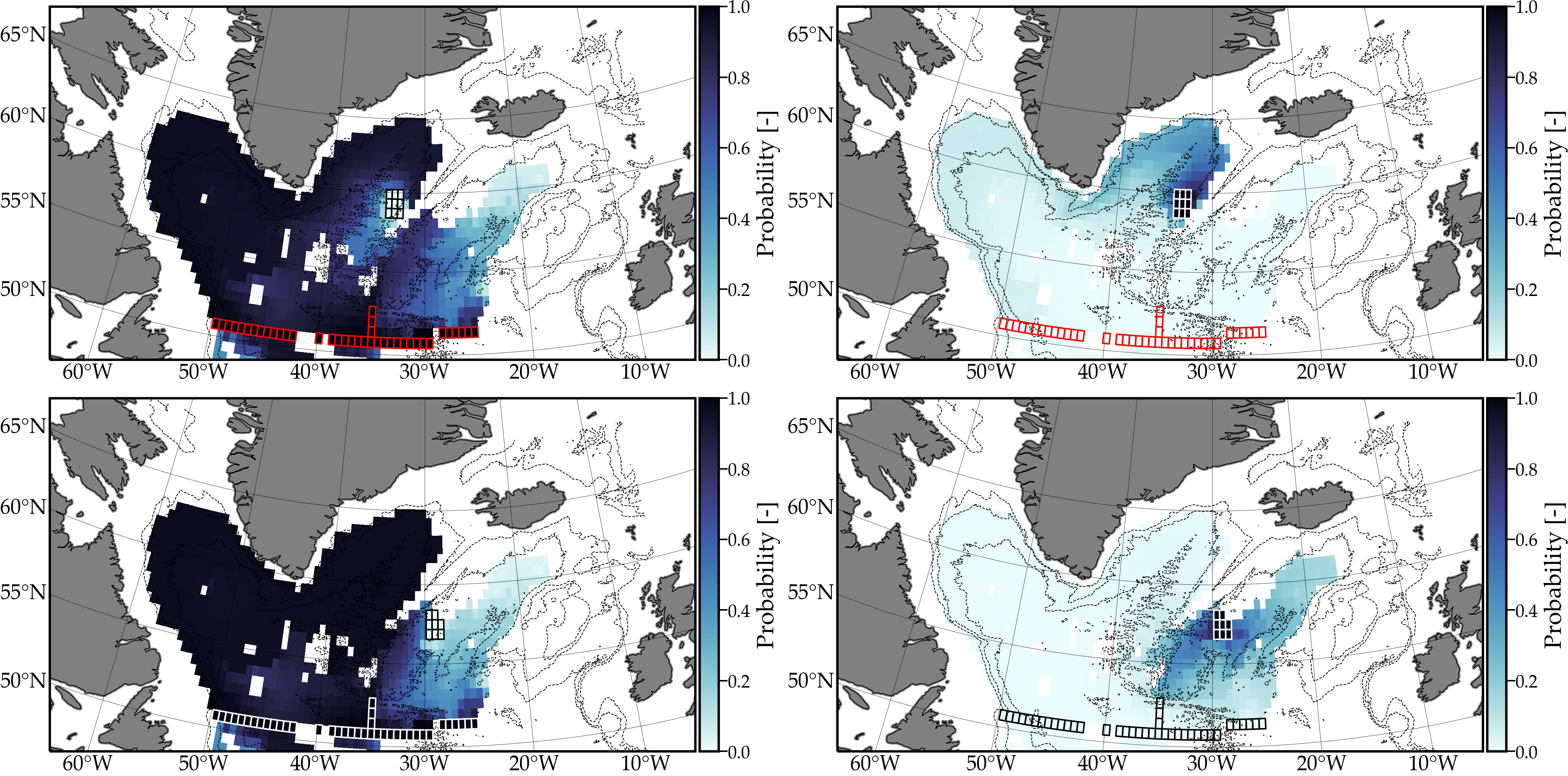} \caption{(top)
  For the LNADW Markov chain model with the source placed west of
  the Reykjanes Ridge, probability of trajectories to commit to the
  target (red boxes) in forward time (left) and to the source
  (black/white boxes) in backward time (right). Reactive trajectories
  (of UNADW) are necessarily small where the forward or backward
  committor probabilities are small. (bottom) As in the top panels,
  but with the source placed east of the Reykjanes Ridge.}
  \label{fig:committor_ow}
\end{figure*}

The left and right panels of Fig.~\ref{fig:committor_ow} show
$\mathbf q^+$ and $\mathbf q^-$, respectively, with the source of
the LNADW chain placed west (top) and east (bottom) of the Reykjanes
Ridge. Similarly to the UNADW chain, $q_i^- \approx 0$ in boxes
covering most of the eastern and the interior of the subpolar North
Atlantic when the source is placed west of the Reykjanes Ridge.
LNADW transition currents out of that source will necessary be
small. When the source is placed east of the Reykjanes Ridge, $q_i^-
\approx 0$ everywhere in the subpolar North Atlantic domain, except
in the West European Basin, where reactive currents can be expected
to concentrate in.

\begin{figure*}[ht!]
  \centering \includegraphics[width=\textwidth]{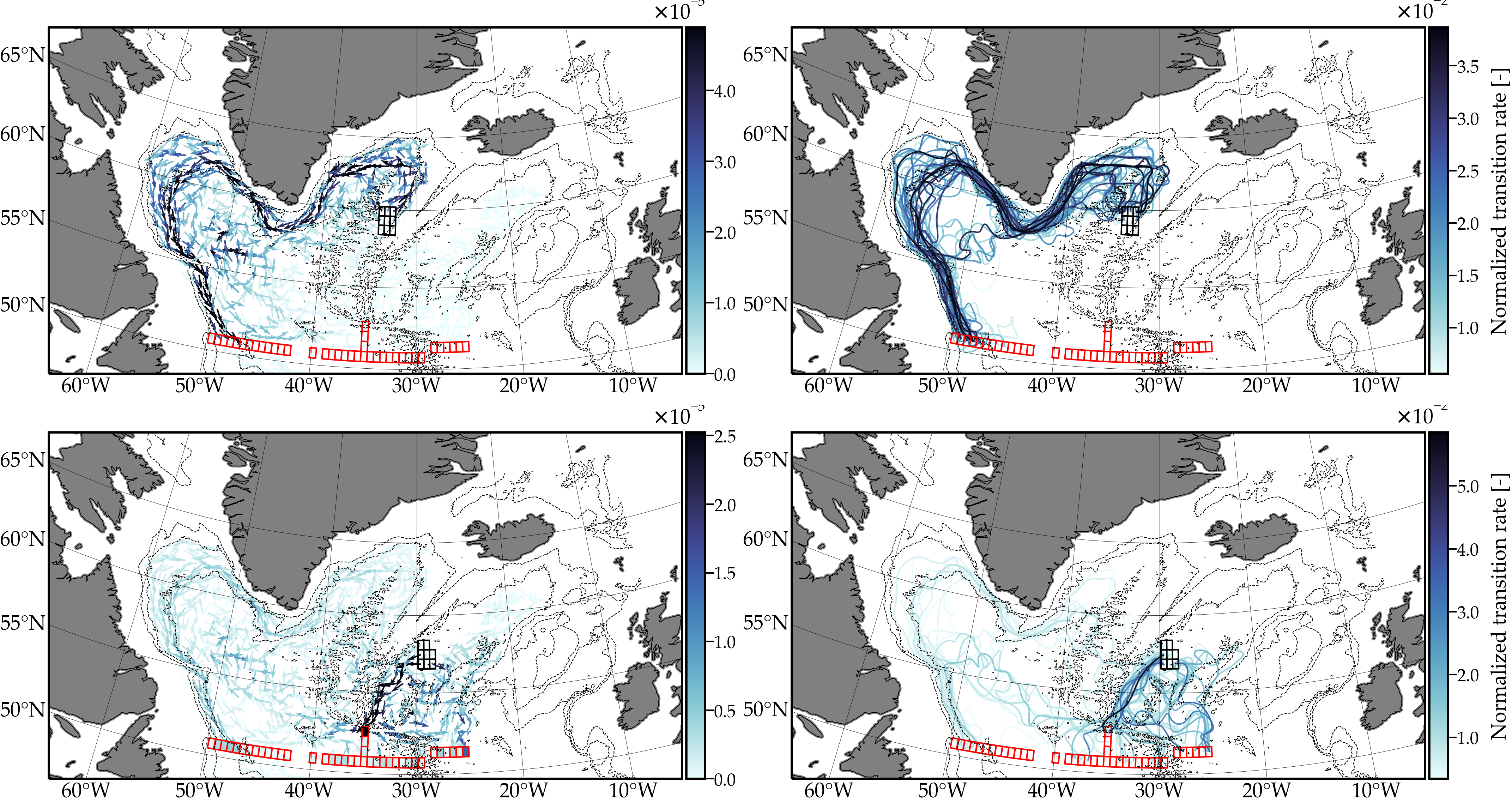} \caption{(top)
  For the LNADW chain with the source west of the Reykjanes Ridge,
  effective reactive currents (left) and dominant transition paths
  accounting for 90\pct{ }of the probability flux of reactive
  trajectories (right). (bottom) As in the top panels, but with the
  source east of the Reykjanes Ridge.} \label{fig:current_ow}
\end{figure*}

The reactive currents (Fig.~\ref{fig:current_ow}, top-left panel)
and dominant transition paths (Fig.~\ref{fig:current_ow}, top-right
panel) of LNADW out of the source west of the Reykjanes Ridge,
reveal a well-defined western DBC, in a manner akin to UNADW reactive
currents and dominant transition paths out of the Irminger Sea
source. The most notable difference is that TPT-inferred DBC in the
Labrador Sea is composed of a single strongly defined branch, which
roughly follows the 3000-m isobath in the Labrador Basin, consistent
with $f/h$ conservation. Note that UNADW transition paths reveal
two branches even if the source is placed in the same location west
of the Reykjanes Ridge used for the TPT analysis of the LNADW chain.

The TPT-inferred path from east of the Reykjanes Ridge, which are
taken to roughly represent the spreading of Iceland--Scotland OW
(Fig.~\ref{fig:current_ow}, bottom panels), differ from the transition
pathways of LNADW sourced west of the Reykjanes Ridge in both
direction and intensity as determined by the probability of reaching
the target, which is smaller. About 70\pct{ }of the reactive currents
are directed southward. Nearly 30\pct{ }of this portion of the reactive
paths flow toward the northward extension of the target at \ang{35}W
and the balance reaches boxes east of it toward the West European
Basin, as anticipated by the analysis of the forward committor
probability (Fig.\@~\ref{fig:committor_ow}, right panels). Some
low-probability transition paths take a westward direction just
before reaching the target around CGFZ and cross the subpolar North
Atlantic interior to cross into the Labrador Sea and rejoin the DBC
in that basin (Fig.~\ref{fig:current_ow}, bottom panels). From the
eastern to the western side of the Reykjanes Ridge, the magnitude
of reactive currents decreases significantly, indicating that very
few transitions connect one side of the ridge with the other. The
results just described are not affected by the meridional portion
of the target.

We note that transition paths for LNADW resemble those for UNADW
when the TPT analysis of LNADW uses UNADW sources. The only difference
if that a two-branch DBC is not so well revealed, likely because
of insufficient data coverage. Transition paths for UNADW with LNADW
sources reveal much more organization, with the development of a
quite well-defined DBC around the Reykjanes Ridge, and eventually
a DBC with the separation into two branches and the subsequent
merging as described above.

\subsection{Mean circulation}

\begin{figure*}[ht!]
  \centering \includegraphics[width=\textwidth]{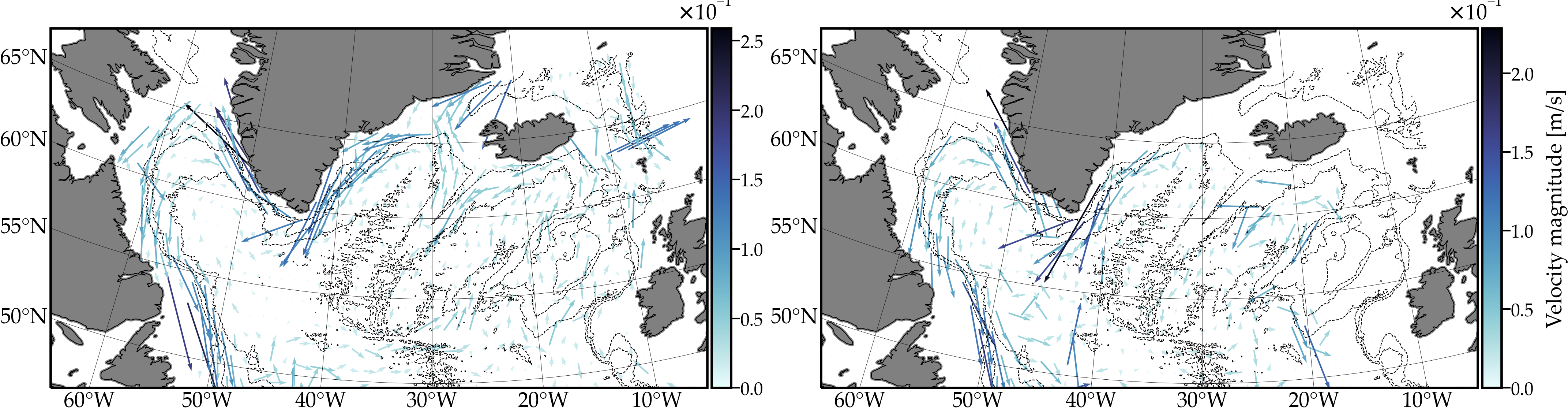} \caption{Mean
  Eulerian velocities obtained by averaging float velocities in
  UNADW (left) and LNADW (right) layers within each box of the
  domain.} \label{fig:eulerian}
\end{figure*}

It is valid to ask if the simple analysis of the mean Eulerian flow
field, could reveal the same characteristics of the NADW paths.
Figure~\ref{fig:eulerian} shows the mean velocity fields obtained
by averaging the float velocities in the UNADW (left) and LNADW
(right) layers within each box of the corresponding domain. Deducing
the pathways previously identified by TPT seems hard, particularly
the interior pathways since the mean Eulerian velocities away from
the western boundary are negligibly small. This is consistent with
the assessment by \citet{Miron2019a}, who analyzed the possibility
that the assumption of time homogeneity of the statistics involved
in constructing their deep-flow Markov chain model, similar to those
underlying the TPT analysis here, could be thought to be represented
simply by advection by the mean circulation with the addition of
(eddy) diffusion. Substantial differences were found, mainly because
their Markov chain model, while time homogeneous, was derived from
trajectories sustained by time-varying deep-ocean currents and
represent, in a statistical sense, the advection--diffusion dynamics
of such a \emph{time-dependent flow}. Here we add that even if the
velocity data were known with infinite resolution, the TPT trajectories
can never be expected to resemble flow \emph{trajectories}, because
they actually highlight the most effective of all such trajectories
connecting predefined source and target locations, i.e., they do
not account for unproductive detours.

\subsection{Reactive rates of UNADW and LNADW}

\begin{figure*}[ht!]
  \centering \includegraphics[width=1\textwidth]{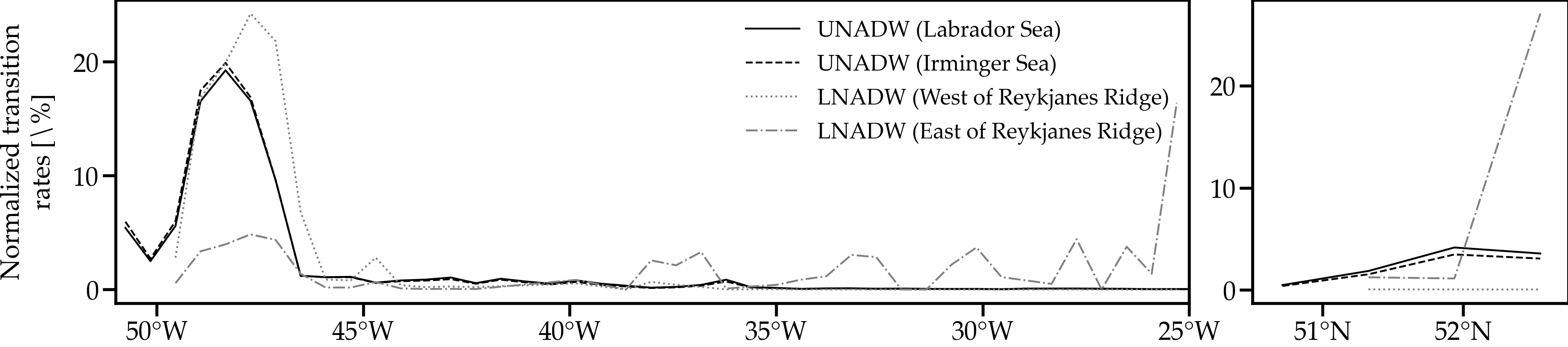} \caption{Reactive
  rates into each latitudinal (left) and meridional (right) box of
  the target(s) normalized by total reactive rate, for UNADW and
  LNADW with the source location as indicated in parenthesis.}
  \label{fig:transport}
\end{figure*}

To further quantify the distribution of transition pathways along
the target, Fig.~\ref{fig:transport} presents reactive rates entering
each target box from the various source, i.e., $k^{b_i\leftarrow}$
for each $b_i \subset B$, normalized by the total reactive rate
into it, i.e., $\smash{k^{B\leftarrow} = \sum_{i\in B} k^{b_i\leftarrow}}$,
for UNADW and LNADW from the various sources considered. The results
are shown as a function of longitude along the zonal piece of the
target (left panel) and of latitude along its meridional piece at
\ang{35}{N} (right panel).

The reactive rates into the latitudinal boxes of the target reveal
a jet-like structure west of \ang{45}W, for UNADW and LNADW
irrespective of the source. In fact, about 79\pct{ }(82\pct) of the
reactive currents of UNADW out of the source in the Labrador
(Irminger) Sea reach the target west of \ang{45}W. For the LNADW,
roughly 94\pct{ }and 19\pct{ }of the currents reach the target west of
\ang{45}W respectively for the LNADW chain sources east and west
of the Reykjanes Ridge.

The reactive rates of UNADW (Labrador and Irminger Seas) present
respectively 11\pct{ }and 10\pct{ }of interior pathways between
\angRange{35}{45}W. For the LNADW with the source west of the
Reykjanes Ridge, internal pathways account for a mere 5\pct{ }of
the direct connections. The story is quite different for LNADW with
the source east of the Reykjanes Ridge, for which more than 70\pct{
}of the pathways to the target materialize outside of the western
boundary current with a large peak around \ang{25}W along with
smaller peaks in several location west of it.

Furthermore, a nonnegligible amount of UNADW reactive rate, about
10\pct{ }from the Labrador Sea source and 8.5\pct from the southwestern
Irminger Sea source, reach the meridional set of boxes of the target
concentrated at \ang{52}N. This is consistent with inferences of
zonal transport of the UNADW by \cite{Goncalves2020}. For LNADW
with the source east of the Reykjanes Ridge, nearly 30\pct reach
the meridional target with a maximum around \ang{52.5}N. This maximum
is most likely due to pathways reaching the area around the CGFZ
but we do not expect important westward crossing through the fracture
zone since eastward transport was shown in \cite{Bower2008} and
more recently in \cite{Goncalves2020}. Nonetheless, \cite{Bower2008}
pointed out that westbound crossings occur north of the CGFZ at
\ang{53}N and more frequently between \angRange{55}{58}N.

These reactive rate calculations are consistent with the results
from the inspection of the reactive currents and dominant paths
that the route of NADW out the subpolar North Atlantic is predominantly
in the form of a well-defined DBC. The UNADW reactive rates into
the meridional target, albeit small, indicate transition paths from
the western to the eastern subpolar North Atlantic through the CGFZ.

\subsection{Dynamical decomposition of the domain}\label{sec:region}

\begin{figure*}[ht!]
\centering
  \includegraphics[width=\textwidth]{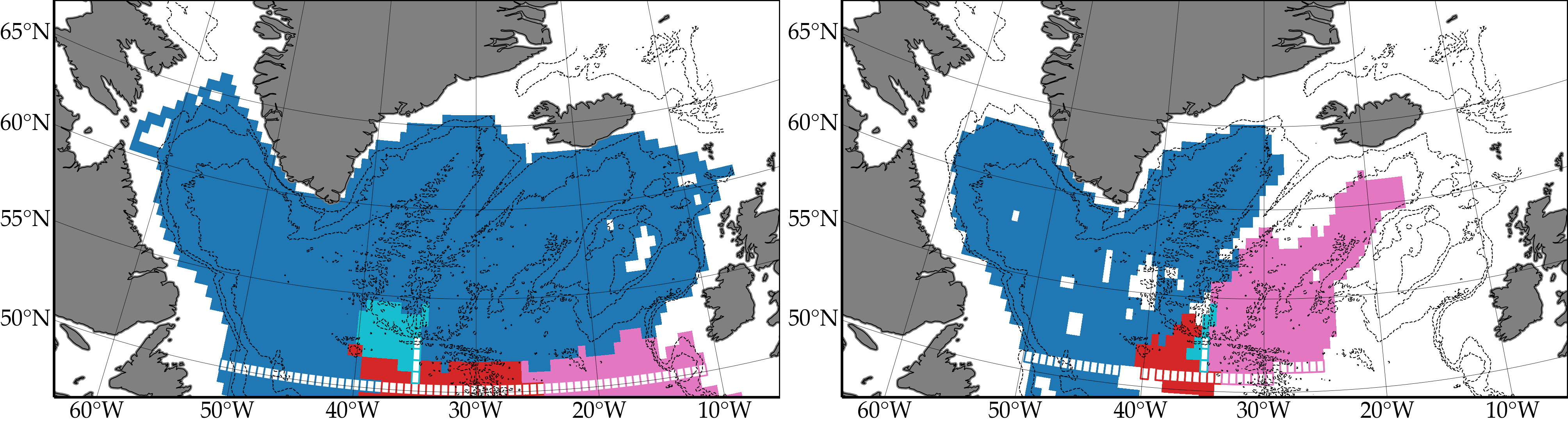} \caption{Decomposition
  of the flow domain according to the maximal probability of reactive
  trajectories sourced from a virtual box outside the flow domain
  to forward committing to each of the four subtargets into which
  the target in the previous figures has been split, both for the
  UNADW (left) and LNADW (right) chains.} \label{fig:regions}
\end{figure*}

The flow domain decomposition in Fig.~\ref{fig:regions}, called a
\emph{forward-committor-based dynamical geography} \citep{Miron2021a},
offers further insight into the NADW Lagrangian dynamics. This is
accomplished as follows.

First, we place the source in the virtual two-way nirvana set
($\omega$) used to close the system \eqref{eq:closure}, in this
case in such a way that probability imbalance is sent back to the
chain uniformly. Since the reactive trajectories emerging from
$\omega$ will necessarily have to travel through the boxes of the
domain on their way to the target ($B$), TPT will unveil transition
paths into $B$ in the southern edge of the flow domain $D$ \emph{without
the restriction of emerging from preset source sets}. Second, we
split the target $B$ at the southern edge of the domain into four
subsets, three of them composed of zonal boxes along it, spanning
nearly equally long longitudinal bands, and the fourth one comprised
of its meridional boxes. Finally, the domain boxes are divided into
four groups according to the maximum probability of each box to
forward committing to each of the four subtargets defined above,
i.e., according to whichever target is most likely to be reached
from a particular grid box. Each group is colored according to the
color assigned to the subtarget to which it most probably forward
commits. The dynamical geography so constructed is formed by provinces
analogous to backward-time basins of attraction \citep{Miron2017,
Miron2019a}, which constrain the transport of UNADW and LNADW into
the corresponding (sub)target.

More precisely, for UNADW (Fig.~\ref{fig:regions}, left panel)
reactive pathways from almost all boxes of the domain converge
toward the westernmost subtarget (blue). The remaining subtargets
(cyan, red, and magenta) are reached by transition channels emerging
from regions in their close vicinity. Unlike the UNADW, which roughly
shows a single dynamical province, the LNADW geography
(Fig.~\ref{fig:regions}, right panel) includes two large provinces
(blue and magenta) separated by the Reykjanes Ridge. The red and
cyan subtargets are seen to present an even smaller influence on
the LNADW reactive flow dynamics than similarly located targets on
that of UNADW. For both UNADW and LNADW, interior easterly reactive
trajectories reach the red and cyan targets, from regions located
west of the Mid-Atlantic Ridge.

\subsection{Expected durations of reactive NADW pathways}\label{sec:timing}

\begin{figure*}[ht!]
  \centering \includegraphics[width=\textwidth]{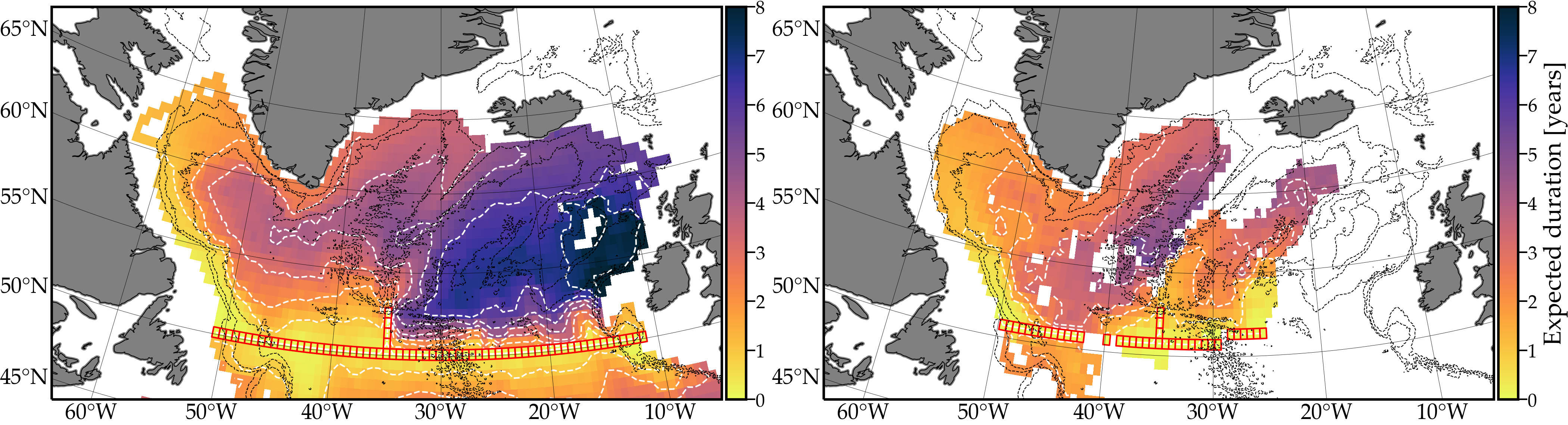} \caption{Mean
  duration of the pathways calculated from boxes of the domain to
  the target (red boxes) located at \angRange{50}{51}N for the UNADW
  (left) and LNADW (right).} \label{fig:hitting_time}
\end{figure*}

Finally, Fig.~\ref{fig:hitting_time} presents expected duration
$\smash{t^{b_iB}}$ of travel of reactive trajectories [cf.\ eq.\
\eqref{eq:t}] out of \emph{each} box $b_i$ to the target $B$ at the
southern edge of the subpolar North Atlantic. As in the sections
prior to the last one, we consider the southern border of the
subpolar North Atlantic as a single target, and transitions are
restricted to happen within the physical flow domain. For UNADW
(Fig.~\ref{fig:hitting_time}, left panel), $\smash{t^{b_iB}}$ is
minimized along the western boundary of the domain (2--3 years),
and clearly very close to the target, where it vanishes. Toward the
east--northeast past the Reykjanes Ridge, $\smash{t^{b_iB}}$
increases, reaching around 8 years. This is suggestive of cyclonic
recirculation and longer direct pathways exiting the subpolar North
Atlantic. For LNADW (Fig.~\ref{fig:hitting_time}, right panel), the
Reykjanes Ridge has a larger impact on the circulation. From the
Iceland Basin, east of the Reykjanes Ridge, the pathways reach the
target about 3--5 years faster than from the Irminger Basin, west
of the Ridge. This is consistent with direct visualization of
multiple paths of RAFOS floats at the southern end of the Reykjanes
Ridge \citep{Zou2017}. Similarly to UNADW, the expected transition
time from the Labrador Sea to the target is estimated between 2--3
years along the western boundary. These results are consistent with
previous pCFC-11 age estimates in the North Atlantic as computed
from observations of CFCs in the North Atlantic \citep{Fine2011}.

\section{Discussion}\label{sec:discussion}

A discussion of the present results in connection with earlier
\citep{Talley1982} and more recent \citep{Lozier2019, Zou2020,
Georgiou2021, Koelling2021} results follows.

Using potential vorticity as a tracer, \citet{Talley1982} have
concluded that the advection of UNADW (Labrador Sea Water) is
accomplished along three main directions: northeastward into the
Irminger Sea, southeastward across the North Atlantic, and southward
along the Labrador Current.  We observed an important recirculation
toward the Irminger basin as well as a dominant southward advection
along a DBC. The northeastward transport of the UNADW is not so
well captured by the TPT analysis, but this may be explained by two
main differences with the analysis of \citet{Talley1982}. First,
here we considered observations in the range 750--1500 m to represent
UNADW, while \citet{Talley1982} considered depths between 500 and
2000 m. At shallower depths, between 100 and 1000 m, inspection of
RAFOS floats by \citet{Bower2008} suggested an eastward flow between
45 and 53$^\circ$N. Second, the target of our TPT analysis is
positioned north of the location where \citet{Talley1982} observed
transport extending to the east towards the Mid-Atlantic Ridge,
which is influenced by the meandering of the northeastward Gulf
Stream Extension/North Atlantic Current.

Our expected duration estimates appear to overestimate the transit
time of deep water in the North Atlantic compared to transit time
estimates by other authors. For example, \citet{Pickart2003} estimated
a transit time out of the subpolar region from the Labrador Sea of
about 1 year. On the other hand, \cite{Georgiou2021} estimated that
transit times between 0.5 and 2 years from the Cape Farewell to
\ang{53}N from Lagrangian particle tracking using global circulation
model output. Since the expected duration represents an average
transit time of reactive trajectories (direct connections between
source and target), the noted differences are not unexpected. Indeed,
the expected duration along the DBC can be anticipated to be shorter,
but interior pathways have been suggested to last within the region
for over 6 years \citep{Georgiou2021}. As pointed out by these
authors, the internal pathways contribute to the variability of the
Atlantic meridional overturning circulation.  Quantify their effects
is beyond the scope of the TPT analysis in its current setting.  A
recent extension thereof, which introduces an explicit probability
flow decomposition into direct transitions and loops supported on
cycles, may provide means for doing it \citep{Banisch2015,
Helfmann2021}.

The seasonal and interannual variabilities suggested by the
measurements taken during the OSNAP Program \citep{Lozier2019}
cannot be quantified by the present analysis, which relies on a
time homogeneity assumption. The extracted pathways of the LNADW
show a higher connection outside of the WBC, as observed by
\cite{Koelling2021}.  This might explain the disconnection pointed
out by \citet{Lozier2019} between water formation from deep water
sources and export outside of the subpolar region. In accordance
with the results presented here, \cite{Koelling2021} also indicate
a longer residence time of internal pathways and a rapid export of
UNADW through the WDBC.

In turn, from the analysis of a subset of the RAFOS trajectories
considered here and the aid of simulated trajectories, \cite{Zou2020}
redrew the pathways of LNADW coming from the CGFZ.  Rather than
turning northward along the western flank of the Reykjanes Ridge,
most of the LANDW was inferred to follow a west-northwestward path
or travel equatorward along the western flank of the Mid-Atlantic
Ridge. A similar picture was drew by \cite{Johns2021} from the
compilation of earlier results by others from float data and
simulations, and from the analysis of moored current meter observations.
The main difference was that they inferred a southward flows along
the eastern flank of the Mid-Atlantic Ridge.  These inferences are
only partially in agreement with our results.  The main agreement
regards the west-northwestward, which TPT confirms.  The major
discrepancies are two. First, TPT does not discard a turning northward
along the western flank of the Reykjanes Ridge (\cite{Johns2021}
does not excludes this path; actually, they suggest one through the
Bight Fracture Zone, which TPT also highlights).  Second, TPT unveils
several equatorward routes.  These include: a nonnegligible WDBC
(produced by the turning northward along the western flank of the
Reykjanes Ridge), several southward routes west of the Mid-Atlantic
Ridge, and a prominent southward path along the eastern flank of
the Mid-Atlantic Ridge (Fig.\ \ref{fig:transport}).

\section{Summary and conclusions}\label{sec:conclusion}

Several aspects of the circulation of North Atlantic Deep Water
(NADW) have been quantified by applying a recent adaptation to open
dynamical systems of Transition Path Theory (TPT) on available RAFOS
and Argo float trajectories. TPT was developed to statistically
characterize ensembles of so-called reactive trajectories. Such
trajectories transition directly from a source to a target, i.e.,
they do not include trajectory detours that unproductively contribute
to transport. We used this characteristic of reactive trajectories
to unveil most effective equatorward routes for NADW within the
subpolar North Atlantic. The modeling framework of TPT analysis is
given by a Markov chain, under the assumption of advection--diffusion
dynamics.

Two Markov chain models were constructed by discretizing the
Lagrangian dynamics as described by the float trajectories. One
chain involved floats at parking depths in the 750--1500-m range
to represent the upper layer of NADW (UNADW), mainly comprised of
Labrador water. The other chain used floats with parking depth
ranging between 1800 and 3500 m, representative of the lower layer
of NADW (LNADW) mainly comprised of Overflow Water. Preset sources
of UNADW in the Labrador and southwestern Irminger Seas, and of
LNADW west and east of the Reykjanes Ridge were considered in the
TPT analysis. The target was located along $50^\circ$N, taken to
represent the southern edge of the subpolar North Atlantic to within
the limits imposed by the availability of data and the depth of the
water masses.

The UNADW component of the TPT-inferred DBC was found to flow between
the 2000- and 3000-m isobaths, while showing two branches in the
Labrador Sea where those isobaths diverge. Each branch follows each
of these isobaths closely. Interior paths of UNADW through the
Charlie--Gibbs Zone were also unveiled, but amounting to a reduced
fraction, not exceeding 20\pct, of the total UNADW transitions. The
majority of the LNADW transition paths (95\pct) tracked out of a
source west of the Reykjanes Ridge were found to form a DBC, well
organized roughly along the 3000-m isobath. LNADW tracked out from
a source east of the Reykjanes Ridge revealed a lesser degree of
organization. About 20\pct{ }of the transitions organized along a
DBC, while nearly 30\pct{ }converged east of the Mid-Atlantic Ridge
on the target at the southern edge of the domain. The rest of the
transitions hit the target in several places distributed along
thereof, indicating southward flow along the western flank of the
Mid-Atlantic Ridge as well. A source-independent dynamical decomposition
of the flow domain into analogous backward-time basins of attraction
consistently reveal a much wider influence of the western side of
the target for UNADW than for LNADW.  The former spans nearly the
entire subpolar North Atlantic, while the latter is confined to the
Labrador and Irminger Seas. For UNADW, the average expected duration
of the pathways from the Labrador and Irminger Seas was found to
range between 2 and 3 years.  For LNADW, the expected duration of
the pathways was found to be largely influenced by the Reykjanes
Ridge.  This found to be as long as 8 years for paths sourced on
the western side of the ridge, while of about 3 years on average
for those on its eastern side.

Future work should aim to expand spatiotemporal coverage and consider
isopycnal ranges, which, unlike fixed depth ranges, most naturally
constrain UNADW and LNADW. Clearly, both are beyond reach of existing
observational platforms, so one will necessarily have to resort to
numerical simulation. At a much more fundamental level lies the
question of the extent to which float motion represents fluid motion,
which can be studied using recent results on the dynamics of inertial
(i.e., buoyant, finite-size) particles \citep{Beron2015}. Addressing
these questions is left for the future.

\section*{Acknowledgments}
We thank Susan Lozier and Kimberly Drouin for their
constructive criticism, and P\'eter Koltai and Luzie Helfmann for
the benefit of many discussions on Transition Path Theory. Support
for this work was provided by the National Science Foundation grant
OCE1851097.

\section*{Data availability statement} 
The RAFOS float data are distributed by the National
Oceanic and Atmospheric Administration's (NOAA) Atlantic Oceanographic
and Meteorological Laboratory (AOML) through the subsurface data
sets website at \url{https://www.aoml.noaa.gov/phod/float_traj/}.
The trajectories of the Argo floats at their parking level are
available in near-real time at
\url{http://apdrc.soest.hawaii.edu/projects/yomaha} \citep{Lebedev2007}.
Finally, the numerical code used in this study is publicly available
at \url{https://github.com/philippemiron/pygtm}.

\bibliographystyle{ametsoc2014}
\bibliography{nadw}

\end{document}